\documentstyle[12pt]{article}

\begin{document}
\begin{titlepage}
\title{On gauge invariance and the path integral }
\author{Silvio J. Rabello\thanks {e-mail: rabello@ntsu1.if.ufrj.br}\,
and Carlos Farina
\\
\\{\it Instituto de F{\'\i}sica}\\
{\it Universidade Federal do Rio de Janeiro}\\
{\it Rio de Janeiro  RJ}\\
{\it Caixa Postal 68.528-CEP 21945-970}\\
{\it Brasil}}
\maketitle
\begin{abstract}

{\sl Using a gauge covariant operator technique we deduce the path integral
for a charged particle in an arbitrary stationary magnetic field, verifying
the ``midpoint rule'' for the discrete form of the interaction term with
the vector potential.}

\end{abstract}

\vfill\hfill\LaTeX
\thispagestyle{empty}
\end{titlepage}

\setcounter{page}{1}
The path integral for a charged particle in a magnetic field gives us a
remarkable example of the subtleties of the discretized time approach and
the relation between path integrals and Brownian motion. When one starts
with the continuum definition of the path integral and tries to perform the
functional integration by time slicing one is faced with the
discretization of the term $\int dt\;{\bf {\dot x\cdot A}}$, that is the
interaction with the vector potential ${\bf A}({\bf x})$, and that for each
infinitesimal segment ${\bf x}_{n+1}-{\bf x}_n$ is given by
$({\bf x}_{n+1}-{\bf x}_n){\bf \cdot A}({\bf x})$.
Naively one could think that it would not matter whether ${\bf A}({\bf x})$
is evaluated at ${\bf x}_n$, ${\bf x}_{n+1}$ or somewhere in between. But it
happens that for the path integral to satisfy the Schr\"odinger equation we
must take the average of ${\bf A}({\bf x})$ between ${\bf x}_n$ and
${\bf x}_{n+1}$ (``midpoint rule''). If we disregard this rule we are
forced to introduce an extra term in the Schr\"odinger equation, breaking
the gauge invariance. The reason for that is in the Brownian character of
the paths in the functional integral \cite{Schu1}. Another approach to
deduce the discretized path integral is to start from the Hamiltonian
formulation of quantum mechanics and define it as an infinite product of
operators where careful use of the Trotter formula  for a proper
``${\bf {\hat x}}-{\bf{\hat p}}$'' ordering leads to the ``midpoint rule''
\cite{Schu2}. In this paper we present an alternative derivation of the
gauge invariant path integral by applying a gauge covariant method developed
long ago by Schwinger to study QED with background fields \cite{Schw}.
In this method instead of the ``${\bf{\hat x}}-{\bf{\hat p}}$'' ordering
we use a time ordering procedure to obtain the configuration space path
integral.

The Hamiltonian operator for a charged particle in an arbitrary stationary
magnetic field is
\begin{equation}
\label{Ham}
 {\hat H}={\hat{\mbox{\boldmath $ \pi$}}^2\over{2m}}\,,
\end{equation}
with ${\hat \pi}^i\equiv {\hat p}^i-e{\hat A}^i$ (i=1,2,3) obeying the
algebra ($\hbar=1$)
\begin{equation}
\label{Fij}
[ {\hat \pi}_i, {\hat \pi}_j]=ie(\partial_i  {\hat A}_j -\partial_j
{\hat A}_i)\equiv ie {\hat F}_{ij}\,,
\end{equation}
and the transition amplitude between the position eigenstates is given by
\begin{equation}
\label{Prop}
\langle {\bf x}'',T\vert {\bf x}',0\rangle=\langle {\bf x}''
\vert e^{-i{\hat H}T}\vert {\bf x}'\rangle\,.
\end{equation}

One way to write a path integral for
$\langle {\bf x}'',T\vert {\bf x}',0\rangle$, is to decompose it
according to Dirac \cite{Di} as

\begin{equation}
\label{Dirac}
\langle {\bf x}'',T\vert {\bf x}',0\rangle=\int d{\bf x}_N\dots
\int d{\bf x}_1 \langle {\bf x}'',T\vert {\bf x}_N,t_N\rangle
\langle{\bf x}_N,t_N\vert{\bf x}_{N-1},t_{N-1}
\rangle\dots \langle {\bf x}_1,t_1\vert {\bf x}',0\rangle\,.
\end{equation}
Then we take $T=(N+1)\varepsilon$ and $t_n=n\varepsilon$ $(n=1,\dots,N)$
letting $N\rightarrow\infty$ and $\varepsilon\rightarrow 0$ with T fixed
\cite {FeyHibbs}. Next we obtain the small time
propagator $\langle {\bf x}_{n+1},t_n+\varepsilon\vert {\bf x}_n,t_n\rangle$,
insert it in (\ref{Dirac}) and try to perform the infinite dimensional
integral.

To evaluate the small time propagator we use a method developed by Schwinger
in his early investigations on effective actions \cite{Schw}. We now write
the transition amplitude as $\langle {\bf x}'',T\vert {\bf x}',0\rangle
\equiv exp(iW)$, where $W({\bf x}'',{\bf x}';T)$ is a complex function of
the end point coordinates and time. Defining the expectation value of an
observable $\hat{\cal O}$ by
\begin{equation}
\label{expect}
\langle {\hat{\cal O}}\rangle \equiv {\langle {\bf x}'',T\vert
{\hat{\cal O}}\vert {\bf x}',0\rangle /{\langle {\bf x}'',T\vert {\bf x}',
0\rangle }},
\end{equation}
it is easy to verify that $W$ is determined by the following equations

\begin{eqnarray}
\label{HJ}
-{\partial W({\bf x}'',{\bf x}';T)\over{\partial T}}
&=&\langle  {\hat H}( {\bf {\hat x}}(T), {\hat{\mbox{\boldmath $\pi$}}}
(T))\rangle ,\\
\bigskip
\label{mom1}
{\partial W({\bf x}'',{\bf x}';T)\over{\partial  x''_i}}&=&\langle
{\hat \pi}_i(T)\rangle +e A_i({\bf x}'') ,\\
\bigskip
\label{mom2}
{\partial W({\bf x}'',{\bf x}';T)\over{\partial x'_i}}
&=&-\langle  {\hat \pi}_i(0) \rangle -e A_i({\bf x}') ,\\
\bigskip
\label{Norm}
W({\bf x}'',{\bf x}';0)&=&-i\, ln \delta^{3} ({\bf x}''-{\bf x}').
\end{eqnarray}

To solve this problem Schwinger noticed that the above
equations relate the transition amplitude to the solution of the Heisenberg
equations for $ {\bf {\hat x}} (T)$ and ${\hat{\mbox{\boldmath $\pi$}}}(T)$.
If we solve for ${\hat{\mbox{\boldmath $\pi$}}}(T)$ in terms of
$ {\bf {\hat x}}(T)$ and $ {\bf{\hat x}}(0)$ and insert this, in a
time ordered fashion, on (\ref{HJ})-(\ref{mom2}) we are left with a set of
first order equations to integrate. For ${\bf {\hat x}}(\varepsilon)$
we have up to second order in $\varepsilon$
\begin{equation}
\label{Heis}
{\hat x}^i(\varepsilon)=e^{i{\hat H}\varepsilon} {\hat x}^i (0)
e^{-i{\hat H}\varepsilon}\simeq {\hat x}^i(0)+{{\hat \pi}^i(0)\over m}
\varepsilon +{e\over 4m^2}(2{\hat F}^{ik}{\hat \pi}_k(0)-i\partial_k
{\hat F}^{ik})\varepsilon^2\,,
\end{equation}
with ${\hat F}^{ik}={\hat F}^{ik}({\bf x}(0))$. Inverting the above equation
to get ${\hat \pi}^i(0)$ in terms of ${\bf {\hat x}}(\varepsilon)$ and
${\bf{\hat x}}(0)$
\begin{equation}
\label{pi0}
{\hat \pi}^i(0)\simeq m{({\hat x}^i(\varepsilon)-{\hat x}^i(0))\over
\varepsilon}-{e\over 2}({\hat x}^k(\varepsilon)-{\hat x}^k(0)){\hat F}^
{ik}\,,
\end{equation}
using the fact that $\langle ({\bf{\hat x}}(\varepsilon)-{\bf{\hat x}}
(0))^2\rangle$ is of order $\varepsilon$ \cite {FeyHibbs}, we see that if
we take $\langle{\hat{\mbox{\boldmath $\pi$}}}(0)\rangle$ the terms in the
above expansion are respectively of order $1/\sqrt{\varepsilon}$ and
$\sqrt{\varepsilon}$. The second  term although small in comparison with
the first, will give a relevant contribution when used in (\ref{mom1}) and
(\ref{mom2}) to evaluate $W({\bf x}_{n+1},{\bf x}_n;\varepsilon)$.
{}From ${\hat{\mbox{\boldmath $\pi$}}}(0)$ we have by time evolution:
\begin{equation}
\label{pie}
{\hat \pi}^i(\varepsilon)\simeq m{({\hat x}^i(\varepsilon)-{\hat x}^i(0))
\over \varepsilon}+{e\over 2}({\hat x}^k(\varepsilon)-{\hat x}^k(0))
{\hat F}^{ik}\,.
\end{equation}
Using the above expression for ${\hat {\mbox{\boldmath $\pi$}}}(0)$ or
${\hat {\mbox{\boldmath $\pi$}}}(\varepsilon)$ in ${\hat H}$,
in a time ordered manner, we are ready to integrate (\ref {HJ})
\begin{equation}
\label{W}
W({\bf x}_{n+1},{\bf x}_n;\varepsilon)\simeq {m({\bf x}_{n+1}-{\bf x}_{n})
^2\over 2\varepsilon} +i{3\over 2} \,ln\,\varepsilon +
\Phi({\bf x}_{n+1},{\bf x}_{n})\,,
\end{equation}
where we used that $[{\hat x}^i(\varepsilon),{\hat x}_i(0)]\simeq
-3i\varepsilon/m$ and $\Phi$ is a time independent function of the end
point coordinates ${\bf x}_{n+1}$ and ${\bf x}_n$. Inserting the above
$W$ in (\ref{mom1}) and (\ref{mom2}) we have for $\Phi$ (remember
$({\bf x}_{n+1}-{\bf x}_n)^2\sim \varepsilon$)

\begin{eqnarray}
\label{phi}
{\partial\Phi \over{\partial x^i_{n+1}}}&=&{e\over 2}
\biggl [\biggl(x_{n+1}-x_{n}\biggr)^k{\partial A_k({\bf x}_{n+1})
\over{\partial x^i_{n+1}}}+ A^i({\bf x}_{n+1})+A^i({\bf x}_n)\biggr ]+
O(\varepsilon)\,,\\
{\partial\Phi \over{\partial x^i_{n}}}&=&{e\over 2}
\biggl [\biggl(x_{n+1}-x_{n}\biggr)^k{\partial A_k({\bf x}_{n})
\over{\partial x^i_{n}}}-\biggl(A^i({\bf x}_{n+1})+A^i({\bf x}_n)\biggr)
\biggr ]+O(\varepsilon)\,,
\end{eqnarray}
with the solution
\begin{equation}
\label{phi2}
\Phi({\bf x}_{n+1},{\bf x}_n)={e\over 2} ({\bf x}_{n+1}-{\bf x}_{n})
\cdot ({\bf A}({\bf x}_{n+1})+{\bf A}({\bf x}_n)) + C\,,
\end{equation}
where C is a constant determined by (\ref{Norm})
\begin{equation}
\label{C}
C=-i{3\over 2}\, ln({m\over{2\pi i}})\,.
\end{equation}

Finally, the small time propagator is
\begin{equation}
\label{Prop2}
\langle {\bf x}_{n+1},t_n+\varepsilon\vert {\bf x}_n,t_n\rangle\simeq
\biggl({m\over{2\pi i\varepsilon}}\biggr)^{3\over 2} e^{iS}\,,
\end{equation}
with
\begin{equation}
\label{S}
S={m({\bf x}_{n+1}-{\bf x}_{n})^2\over 2\varepsilon}+
{e\over 2} ({\bf x}_{n+1}-{\bf x}_{n})\cdot ({\bf A}({\bf x}_{n+1})
+{\bf A}({\bf x}_n))
\,,
\end{equation}
that is precisely the midpoint expansion rule.

We see that by applying a gauge covariant method we were able to obtain the
the ``midpoint rule'' in a natural way.
In our deduction two facts were essential, first the Brownian like relation
$({\bf x}_{n+1}-{\bf x}_n)^2\sim \varepsilon$ and second the gauge
covariance conditions (\ref {mom1}) and (\ref {mom2}). The authors are
grateful to the CNPq (Brazilian Research Council) for the financial support.

\end{document}